\documentclass[iop]{emulateapj}

\usepackage{graphicx}
\usepackage{amsmath}
\usepackage{natbib}
\usepackage{mathptmx}
\usepackage{apjfonts}
\usepackage{times}
\usepackage{multirow}
\usepackage[colorlinks=true,urlcolor=blue,citecolor=blue,linkcolor=blue]{hyperref}
\usepackage{aas_macros}
\usepackage{makecell}

\bibpunct{(}{)}{;}{a}{}{,}

\usepackage[position=t,singlelinecheck=on,font={rm,bf,up},font=large]{subfig}

\makeatletter
\long\def\frontmatter@title@above{
  \vspace*{-17mm}\vspace*{\headheight}
   \hspace{-3mm}{\sc The Astrophysical Journal Supplement Series}, 229:8 (8pp), 2017\\
   \vspace*{4mm}{\footnotesize {\sc Preprint typeset using \LaTeX\ style emulateapj}}
  \par\vspace*{-\baselineskip}\vspace{6mm}
  }

\makeatother

\RequirePackage{color}

\shorttitle{Kinematics of magnetic bright features in the solar photosphere}
\shortauthors{Jafarzadeh et al.}

\begin{document}

\title{Kinematics of Magnetic Bright Features in the Solar Photosphere}

\author{S.~Jafarzadeh\hyperlink{}{\altaffilmark{1}}}
\author{S.~K.~Solanki\hyperlink{}{\altaffilmark{2,3}}}
\author{R.~H.~Cameron\hyperlink{}{\altaffilmark{2}}}
\author{P.~Barthol\hyperlink{}{\altaffilmark{2}}}
\author{J.~Blanco~Rodr\'{i}guez\hyperlink{}{\altaffilmark{4}}}
\author{ J.~C.~del~Toro~Iniesta\hyperlink{}{\altaffilmark{5}}}
\author{A.~Gandorfer\hyperlink{}{\altaffilmark{2}}}
\author{L.~Gizon\hyperlink{}{\altaffilmark{2,6}}}
\author{J.~Hirzberger\hyperlink{}{\altaffilmark{2}}}
\author{M.~Kn\"{o}lker\hyperlink{}{\altaffilmark{7,}\altaffilmark{10}}}
\author{V.~Mart\'{i}nez~Pillet\hyperlink{}{\altaffilmark{8}}}
\author{D.~Orozco~Su\'{a}rez\hyperlink{}{\altaffilmark{5}}}
\author{T.~L.~Riethm\"{u}ller\hyperlink{}{\altaffilmark{2}}}
\author{W.~Schmidt\hyperlink{}{\altaffilmark{9}}}
\author{M.~van~Noort\hyperlink{}{\altaffilmark{2}}}

\affil{\altaffilmark{1}\hspace{0.2em}Institute of Theoretical Astrophysics, University of Oslo, P.O. Box 1029 Blindern, N-0315 Oslo, Norway; \href{mailto:shahin.jafarzadeh@astro.uio.no}{shahin.jafarzadeh@astro.uio.no}\\
\altaffilmark{2}\hspace{0.2em}Max Planck Institute for Solar System Research, Justus-von-Liebig-Weg 3, 37077 G\"{o}ttingen, Germany\\
\altaffilmark{3}\hspace{0.2em}School of Space Research, Kyung Hee University, Yongin, Gyeonggi 446-701, Republic of Korea\\
\altaffilmark{4}\hspace{0.2em}Grupo de Astronom\'{\i}a y Ciencias del Espacio, Universidad de Valencia, 46980 Paterna, Valencia, Spain\\
\altaffilmark{5}\hspace{0.2em}Instituto de Astrof\'{i}sica de Andaluc\'{i}a (CSIC), Apartado de Correos 3004, E-18080 Granada, Spain\\
\altaffilmark{6}\hspace{0.2em}Institut f\"ur Astrophysik, Georg-August-Universit\"at G\"ottingen, Friedrich-Hund-Platz 1, 37077 G\"ottingen, Germany\\
\altaffilmark{7}\hspace{0.2em}High Altitude Observatory, National Center for Atmospheric Research, \footnote{The National Center for Atmospheric Research is sponsored by the National Science Foundation.} P.O. Box 3000, Boulder, CO 80307-3000, USA\\
\altaffilmark{8}\hspace{0.2em}National Solar Observatory, 3665 Discovery Drive, Boulder, CO 80303, USA\\
\altaffilmark{9}\hspace{0.2em}Kiepenheuer-Institut f\"{u}r Sonnenphysik, Sch\"{o}neckstr. 6, D-79104 Freiburg, Germany}

\altaffiltext{10}{The National Center for Atmospheric Research is sponsored by the National Science Foundation.}

\begin{abstract}
Convective flows are known as the prime means of transporting magnetic fields on the solar surface. Thus, small magnetic structures are good tracers of the turbulent flows. We study the migration and dispersal of magnetic bright features (MBFs) in intergranular areas observed at high spatial resolution with {\sc Sunrise}/IMaX. We describe the flux dispersal of individual MBFs as a diffusion process whose parameters are computed for various areas in the quiet Sun and the vicinity of active regions from seeing-free data. We find that magnetic concentrations are best described as random walkers close to network areas (diffusion index, $\gamma=1.0$), travelers with constant speeds over a supergranule ($\gamma=1.9-2.0$), and decelerating movers in the vicinity of flux emergence and/or within active regions ($\gamma=1.4-1.5$). The three types of regions host MBFs with mean diffusion coefficients of $130$~km$^2$\,s$^{-1}$, $80-90$~km$^2$\,s$^{-1}$, and $25-70$~km$^2$\,s$^{-1}$, respectively. The MBFs in these three types of regions are found to display a distinct kinematic behavior at a confidence level in excess of 95\%.

\vspace{2mm}
\end{abstract}


\section{Introduction}
\label{sec:intro}

The kinematics of magnetic structures play an essential role in heating the upper solar atmosphere, e.g., by generating magnetohydrodynamic (MHD) waves~\citep{Jafarzadeh2013a,Jafarzadeh2017a}, or by braiding the field lines through the non-oscillatory motions of their footpoints~\citep{Parker1972,Parker1983,Parker1988}. Both of these processes are produced by interactions between magnetic flux tubes with their surrounding plasma and their characteristics vary in different solar regions. In addition, various magnetic environments have been shown to strongly influence the embedded flows depending on their level of magnetic flux~\citep{Ji2016}. 
In the quiet-Sun internetwork, advection of flux concentrations is described as a superposition of a random component, caused by intergranular turbulence and granular evolution, on a systematic drift due to the large-scale motions of granules, mesogranules, and supergranules~\citep[e.g.,][]{Manso-Sainz2011,Jafarzadeh2014a}. In network areas, the oppositely directed inflows from neighboring supergranules appear to trap magnetic elements in sinks, so that they are expected not to move freely~\citep{Orozco2012c}.

The motion of magnetic features is often described by diffusion processes whose ``index'' and ``coefficient'' (see below for definitions) provide information on how fast a structure moves from its initial position and on the rate of increase in area that a feature sweeps in time, respectively. Hence, the larger a diffusion index ($\gamma$) is, the faster the magnetic element moves away, so that the magnetic flux is spread to a larger extent by the turbulent flows. \citet{Del-Moro2015}, however, claimed that the diffusion parameters, determined from the displacement of magnetic elements, may not necessarily correspond to a turbulent regime.

Magnitudes of diffusion parameters describe various diffusivity regimes \citep[e.g.,][]{Abramenko2011,Jafarzadeh2014a} such as (1) $\gamma=1$, the so-called normal diffusion, where magnetic elements are randomly advected around (random walkers) and the diffusion coefficient ($D$) is independent of temporal and spatial scales; (2) $\gamma<1$, the sub-diffusive process, where features are trapped at so-called stagnation points (sinks of flow field) and $D$ is anti-correlated with both time and length scales; and (3) $\gamma>1$, the super-diffusive case, indicating regions where small structures quickly move away from their first location ($D$ grows with scales both in time and in length). Magnetic elements in the latter regime are transported with a negative acceleration when $\gamma<2$, with a constant average speed for $\gamma=2$ (known as ``ballistic'' diffusion), and with a positive acceleration when $\gamma>2$ (the super-ballistic branch). We note that these characteristics describe the flow field displacing the magnetic features horizontally rather than the structure of magnetic elements. 

Moreover, diffusion coefficients in the solar atmosphere have been shown to be inversely related to the size and field strength of magnetic concentrations~\citep{Schrijver1989,Schrijver1996}. Thus, different magnetic environments on the solar surface, hosting a variety of magnetic features with a variety of properties~\citep{Borrero2015}, may represent different diffusivity behavior.

Most of the previous measurements of diffusion parameters are either focused on magnetic elements in the quiet-Sun or are based on relatively low spatial/temporal resolution observations. It is only recently that meter-class telescopes (such as NST at the Big Bear Solar Observatory~(\citealt{Goode2010}), the Swedish Solar Telescope (\citealt{Scharmer2003}), the broad-band imager on GREGOR (\citealt{Schmidt2012}), and the {\sc Sunrise} balloon-borne solar observatory (\citealt{Solanki2010})) have provided us such information at high spatial resolution. However, only values from the last mentioned observatory are not affected by differential seeing-induced deformations that consequently introduce artificial turbulence in a time series of solar images. \citet{Jafarzadeh2014a} studied diffusivity of magnetic bright points observed in the Ca~{\sc ii}~H passband of the {\sc Sunrise} telescope. Their study was, however, limited to the quiet-Sun and to observations sampling heights corresponding to the temperature minimum and/or low chromosphere. For a review of diffusion parameters of small magnetic elements in the literature, we refer the reader to \citet{Jafarzadeh2014a} (hereafter Paper~I), who also summarized some of those values in Table~3 of their paper.

In the present study, we aim to characterize the statistical properties of the proper motion of individual trajectories of magnetic bright features (MBFs), which is necessary in order to determine whether the action of the flow on the magnetic structures can be interpreted as a turbulent diffusivity and how the action of the flow depends on the amounts of magnetic flux harbored in different regions. To this end, we borrow some of the language of turbulent diffusivity in order to characterize the individual trajectories. Thus, we determine diffusion parameters for individual MBFs in various solar regions with different levels of magnetic activity. We use seeing-free observations with high spatial and temporal resolution obtained with the {\sc Sunrise} balloon-borne observatory (Section~\ref{sec:obs}). We analyze trajectories of MBFs in areas with different amounts of magnetic flux and different types of features (Section~\ref{analysis}) and discuss their diffusion parameters to describe their various plasma environments (Section~\ref{conclusions}). 

\vspace{3mm}
\section{Observational Data}
\label{sec:obs}

Our analysis is based on two data sets recorded with the Imaging Magnetograph eXperiment (IMaX;~\citealt{Martinez-Pillet2011}) on board the {\sc Sunrise} balloon-borne solar observatory~\citep{Barthol2011,Berkefeld2011,Gandorfer2011}, from its first and second flights in 2009 and 2013 (hereafter, {\sc Sunrise}-I and {\sc Sunrise}-II, respectively \citealt{Solanki2010,Solanki2017}). 

The {\sc Sunrise}/IMaX, which is a Fabry--P\'{e}rot based instrument, recorded the full Stokes vector ($I$, $Q$, $U$, and $V$) of the magnetically sensitive line Fe~{\sc i}~$5250.2$~\AA\ (with a single-wavelength noise level of $\approx3\times10^{-3}$ in the unit of the Stokes $I$ continuum, after phase-diversity reconstruction during the 2009 flight). The images obtained during the 2009 flight, on average, cover a field-of-view (FOV) of ($45\times45$)~arcsec$^2$ on the solar surface with a scale of $\approx0.0545$~arcsec/pixel, while 2013 images have a useful FOV of ($51\times51$)~arcsec$^2$.

The data acquired on 2009 June 9 (between 01:32 and 01:58 UT) samples a quiet-Sun area close to disk center, including both network and internetwork regions. The time series of images were obtained with a rate of 33~s and the Stokes parameters were recorded in five wavelength positions at $\pm 40, \pm 80$, and $+227$~m\AA\ from the line center, of which the latter represents a continuum image (see \citealt{Jafarzadeh2014b} for the formation heights).

The second dataset, from {\sc Sunrise}-II (obtained on 2013 June 12; 23:39-23:55 UT) provided us with high-resolution observations of an active region close to the solar disk center (AR 11768 at a heliocentric angle $\mu$=0.93) that includes a small quiet area, a flux-emergence region, diverse pores, and plage areas. This image sequence has a cadence of 36.5~s and samples the Fe~{\sc i}~$5250.2$~\AA\ line at seven wavelength positions inside the line (at 0, $\pm 40, \pm 80, \pm 120$~m\AA\ from the line center) and one at $+227$~m\AA\ in the continuum.

The Stokes $I$ continuum images are the primary data sets for the present study, in which the motion of small MBFs near the base of the photosphere is investigated. In addition, we use the ``line-core'' images and circular polarization (CP) maps to facilitate identification of MBFs and to inspect their magnetic nature, respectively. 
Since we have no observations at the line center of Fe~{\sc i}~$5250.2$~\AA\ from the 2009 flight, we form the line core by averaging the two closest wavelength positions around the line center (which is a combination of the line core and the line's inner flanks). These line-core images (with a relatively large contrast) are used to ease detection of MBF in Stokes $I$ continuum images.
Maps of circular polarization, or CP, are formed as an unsigned average of four inline positions in Stokes $V$ (the closest two positions at each side of the line center; at $\pm80$ and $\pm40$~m\AA\ from the center of Fe~{\sc i}~$5250.2$~\AA\ line), essentially as described in \citet{Jafarzadeh2014b}. The latter integration increases the signal-to-noise ratio (S/N) by approximately a factor of two.

Our intention is to investigate the dispersal of MBFs in various solar regions with different levels and types of magnetic activity, that are provided by the two flights of the {\sc Sunrise} observatory.

\begin{figure*}[!tb]
\centering
    \includegraphics[width=.98\textwidth, trim = 0 0 0 0, clip]{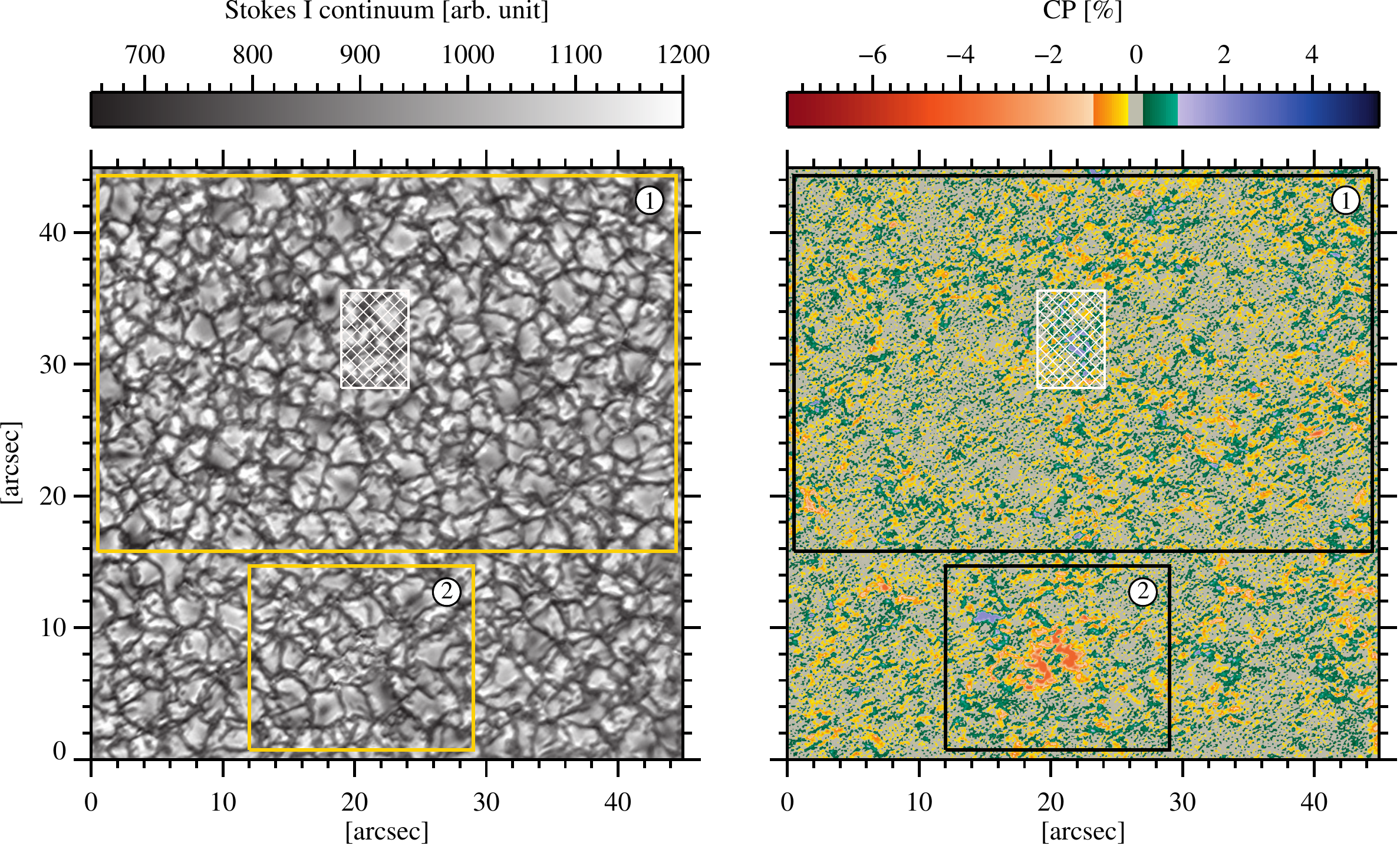}
  \caption{Examples of {\sc Sunrise}/IMaX Stokes $I$ continuum (left) and averaged Stokes $V$/$I_{c}$ (CP; right) images from observations in 2009. The rectangles outline the two regions of interest (ROI) in this time series: an internetwork (ROI-1) and a network (ROI-2) area. The cross-hatched rectangle indicates a network patch excluded from ROI-1.}
  \label{fig:obsI}
\end{figure*}

\begin{figure*}[!tp]
\centering
    \includegraphics[width=.98\textwidth, trim = 0 0 0 0, clip]{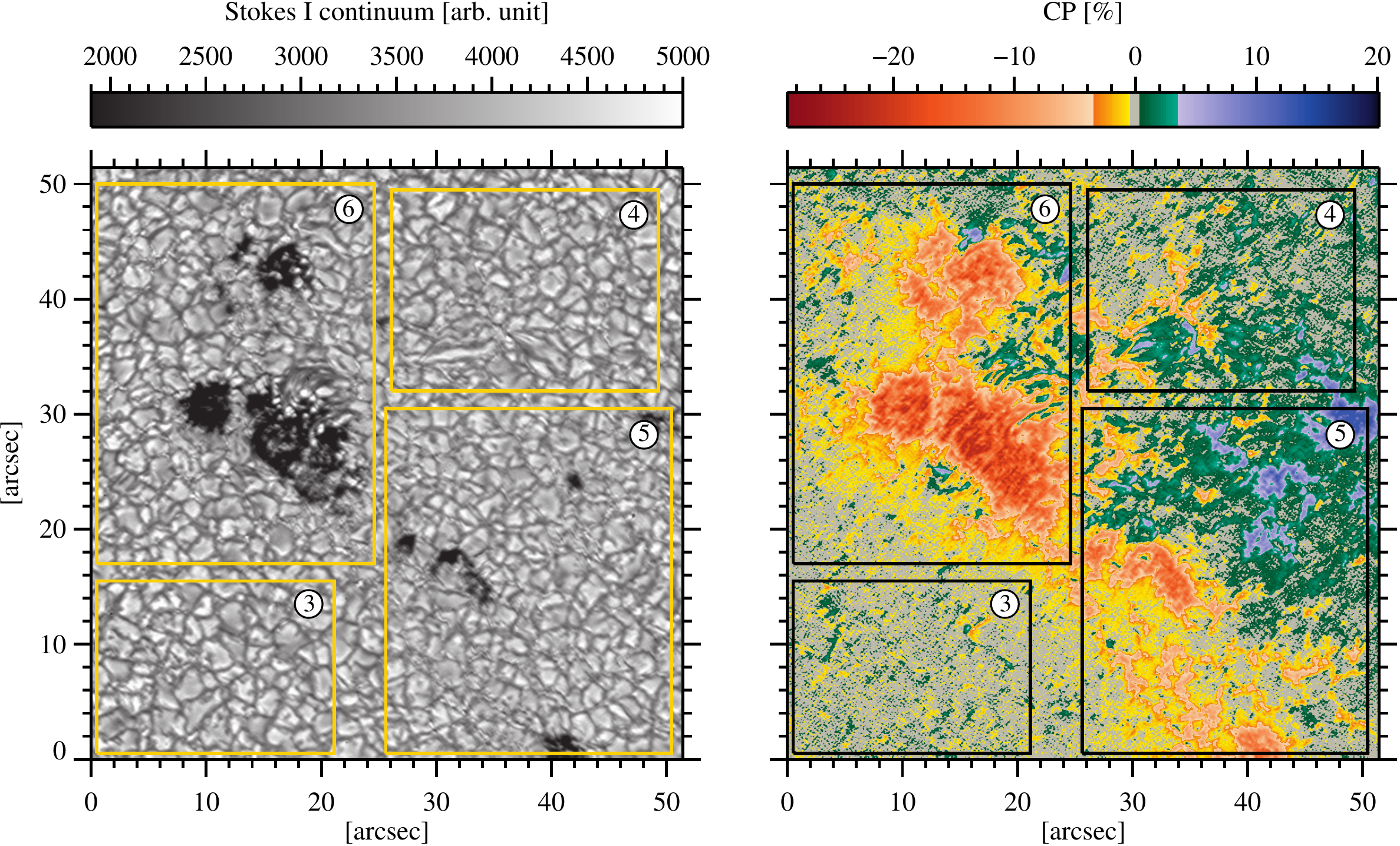}
  \caption{Same as Figure~\ref{fig:obsI}, but made from observations in 2013. The rectangles mark four ROIs, including a quiet-Sun internetwork (ROI-3), an area with flux emergence and plages (ROI-4), small pores and plage (ROI-5), and a region with large pores and a high magnetic flux density (ROI-6).}
  \label{fig:obsII}
\end{figure*}

Figure~\ref{fig:obsI} illustrates an example of a Stokes $I$ continuum image from {\sc Sunrise}-I (left panel) along with its corresponding CP map (right panel). Two regions of interest (ROI) are indicated by the rectangles (along with their labels), representing a quiet-Sun internetwork (ROI-1) and a quiet-Sun network area (ROI-2). A small network patch close to the center of ROI-1 (marked by a crossed-hatched area) is excluded from the internetwork region.
Examples of Stokes $I$ continuum and CP images from {\sc Sunrise}-II are shown in the left and in the right panels of Figure~\ref{fig:obsII}, respectively. Here, we have visually selected four ROIs with different (although sometimes similar) levels of magnetic activity. These regions include an internetwork area (ROI-3), a flux-emergence region (ROI-4), a few small pores including plages (ROI-5), and an area with relatively large pores and an evolving sunspot (ROI-6). We note that the FOV of the images in Figure~\ref{fig:obsII} is vertically flipped and slightly rotated with respect to the true orientation on the Sun. For the correct orientation, see \citet{Solanki2017}.

\vspace{3mm}
\section{Analysis and Results}
\label{analysis}

We characterize trajectories of MBFs on the solar surface by means of a diffusion analysis. The trajectories are formed by linking locations of MBFs in a time series of images. We detect MBFs and determine their precise locations using the same procedure as described in \citet{Jafarzadeh2014b}. We perform the detection algorithm simultaneously on both Stokes $I$ continuum and Stokes $I$ line-core images, the latter being a better guide for visual identification of the MBFs (because of a larger intensity contrast compared to the continuum). An MBF in our analysis is defined as a bright structure (i.e., with an intensity contrast larger than the average quiet-Sun) residing in intergranular areas if (1) it is magnetic (i.e., it coincides with $CP\geqslant3\sigma_{noise}$), (2) it lives for 80~s or longer, and (3) it has a roughly circular shape (to avoid apparently connected or merged magnetic elements), and does not show interactions with other magnetic features (including merging and/or splitting) during the course of its lifetime. Unlike in Paper~I, we are not setting a size threshold on MBFs. However, features with a strongly non-uniform brightness structures are excluded during our visual identification. The latter structures have been shown to increase uncertainty in measuring the locations of features.

The location of MBFs is determined with an accuracy of 0.05 pixel, i.e., 2~km. In accordance with a discussion by \citet{Jafarzadeh2014b}, we, however, consider a more conservative uncertainty of 0.5 pixel (19~km), which takes the effects of temporal size and intensity variation of the features into account. The detected features are then tracked in image sequences using the approach introduced by \citet{Jafarzadeh2013a}.

\subsection{Diffusion Analysis}
\label{diffusion}

Magnetic features in the solar photosphere, and in particular our detected MBFs, can be considered as ``fluid particles'' in a Lagrangian approach, transported by turbulent flows. In the Lagrangian method, the particles' velocities are determined by tracking and together form a velocity field. Analysis of the velocity field provides statistical properties of the flow in which the particles are embedded~\citep{Monin1975}.
According to this approach, the Lagrangian form of the diffusion process is described as $\langle sd\rangle\propto\tau^{\gamma}$, where the exponent $\gamma$ represents the diffusion index and $\langle sd\rangle$ is the mean square displacement of all features from their initial locations at time $\tau$.

The mean square displacement cannot, however, provide full information about the dynamics of a system that potentially accommodates more than one type of motion, hence, it does not identify them clearly as separate processes (see, e.g., \citealt{Dybiec2009}).
From a simple simulation in Paper~I, it was shown that an MBF in an internetwork area may have motions with positive, zero, or negative acceleration, depending on its location on a supergranule with respect to the supergranular boundaries.
Hence, it is essential to separately perform the diffusion analysis on individual trajectories, which can provide proper insight into their motion characteristics. For recent papers based on the latter approach (i.e., diffusion index of individual MBFs) see, e.g., \citet{Jafarzadeh2014a}, \citet{Keys2014}, and \citet{Yang2015}. Following \citet{Yang2015}, we call this method distribution of diffusion indices (DDI) henceforth.

Since the present work is aimed at characterizing the statistical properties of the individual trajectories, we use the DDI approach, which borrows some of the language of turbulent diffusivity.
Thus, the diffusion process of individual tracers in the present study is simply described by 

\begin{equation}
	sd(\tau)=C \; \tau^{\gamma}\,,
	\label{equ:index}
\end{equation}

\noindent
where $C$ is the constant of proportionality from which the diffusion coefficient ($D$) is calculated 

\begin{equation}
	D(\tau)= \frac{\gamma}{4\tau} \; C \; \tau^\gamma\,.
	\label{equ:coefficient}
\end{equation}

In practice, we measure $\gamma$ and $C$ from the slope and $y$-intercept of the least-squares fit to the log-log plot of the $sd(\tau)$ of individual MBFs, respectively. Thus the lifetime of each MBF is considered as the timescale $\tau$. For further details of our diffusion analysis as well as examples of various diffusion regimes, we refer the reader to Section~3.2 of Paper~I.

We note that the difference between diffusion parameters calculated using the Lagrangian method (i.e., with $\gamma$ determined from the average of displacements of individual MBFs) and the DDI method (which is based on the mean of diffusion indices of individual MBFs) has been extensively compared and discussed by \citet{Yang2015}, who have shown that the diffusion parameters resulting from the two methods are similar, although slightly different. \citet{Yang2015} showed that the Lagrangian approach results in a smaller $\gamma$, but a larger $D$ compared to those calculated from the DDI method. They found that the differences of the $\gamma$ and $D$ values from the two approaches were smaller than 20\% and 30\%, respectively.
A comparison between diffusion indices of several G-band bright points (GBPs) with different characteristics has been provided by \citet{Yang2015b} by over-plotting $sd(\tau)$ of the GBPs as well as their mean square displacement versus $\tau$ in the same figure.

\subsection{Results and Statistics}
\label{results_obs}

\begin{figure*}[!htp]
\centering
    \includegraphics[width=.96\textwidth, trim = 0 0 0 0, clip]{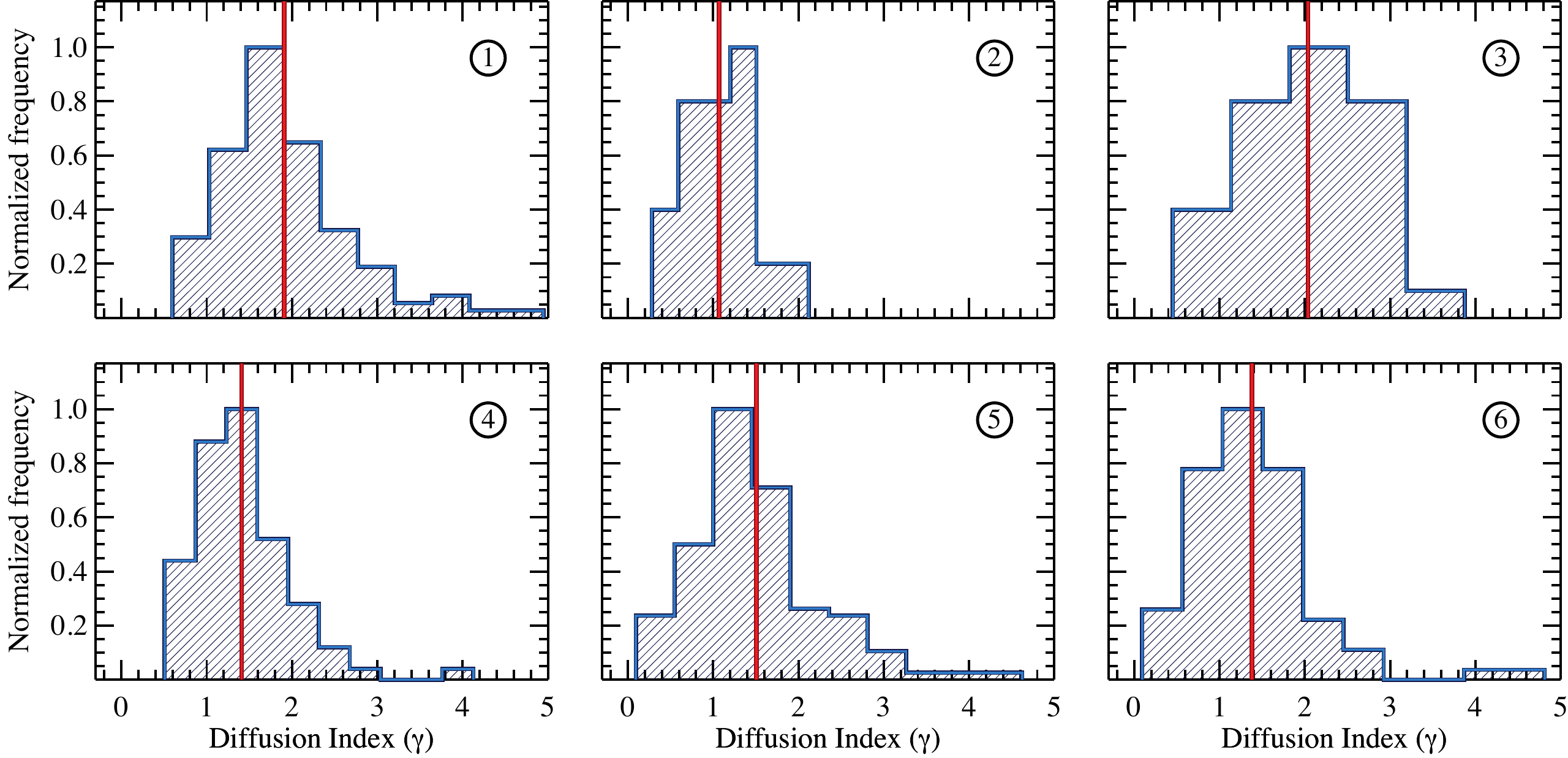}
  \caption{Distributions of diffusion index of magnetic bright features observed in the six regions of interest from {\sc Sunrise}/IMaX, here labeled 1--6 in accordance with the labeling given to them in Figures~\ref{fig:obsI} and \ref{fig:obsII}. The vertical lines indicate mean values of the histograms. The histograms are normalized to their maximum values.}
  \label{fig:stat_obs}
\end{figure*}

Figures~\ref{fig:stat_obs}(1)--(6) are plots of distributions of the diffusion index of the six different solar regions marked in Figures~\ref{fig:obsI} and \ref{fig:obsII}. The label in the top-right corner of each panel corresponds to the number of the ROI indicated in Figures~\ref{fig:obsI} and \ref{fig:obsII}.

We note that, due to the relatively short length of the IMaX time series (particularly for observations from {\sc Sunrise}-II) as well as the relatively small areas of the ROIs, we have detected a relatively small number of MBFs in each region. Table~\ref{table:stat_obs} summarizes the number of detected features along with their diffusion parameters for the six different ROIs. 

As we show below, the reliability of our results is not affected by the relatively small number of MBFs in each region. Firstly, our well-developed semiautomatic algorithm has ensured true detection and precise tracking of the features under study, which results in adequate distributions of diffusion indices. Secondly, we performed Student's \textit{t}-test which determines if two sample populations (with or without equal variances) are significantly different from each other \citep[e.g.,][]{Yuen1973,Yuen1974}. The result of this statistical test confirmed (with a confidence level better than 95\%) that the three groups of MBFs located in internetwork, network, and active regions (see Table~\ref{table:stat_obs}) are indeed different, while  the groups of MBFs in regions with similar magnetic activity belong to the same kinematic population. This means that MBFs in the various internetwork areas (ROI-1 and ROI-3), or active region areas (ROI-4, ROI-5, and ROI-6), display a similar kinematic behavior.

The relatively wide ranges of the distributions of $\gamma$ have been shown to be a result of realization noise (due to the short lifetimes of the features under study; see Paper~I). Hence, the mean values of the diffusion indices are considered to describe the diffusion properties in a particular region relatively well, while the dispersion of the individual diffusion parameters is not considered to be of much diagnostic value for the diffusion properties.

Our analysis of the {\sc Sunrise}/IMaX data reveals that the diffusion index is strongly dependent on the magnetic environment in which the MBFs under study reside. Thus, active regions host MBFs whose motions are best described by a super-diffusive, but sub-ballistic, regime ($\gamma\approx1.4--1.5$). The latter correspond to a decelerating motion, according to the $sd(\tau)\propto\tau^{\gamma}$ relationship. The MBFs within supergranules (i.e., in internetwork areas) are found to have a diffusion index on the order of 2, meaning motions with roughly constant velocities. Finally, features detected around the network patch in ROI-2 (in Figure~\ref{fig:obsI}) are best described as random walkers (i.e., normal diffusion; $\gamma\approx1$).

The diffusion coefficients, as summarized in Table~\ref{table:stat_obs}, have the smallest values for MBFs detected in the active region ($D\approx25-70$~km$^2$\,s$^{-1}$). Interestingly, these values decrease with the level of activity, having the smallest value of 25~km$^2$\,s$^{-1}$ in ROI-6 around the big pore (see Figure~\ref{fig:obsII}), a mean value of 40~km$^2$\,s$^{-1}$ around the small pores in ROI-5, and the largest values compared to the other two ROIs in the active region in ROI-4 ($D\approx70$~km$^2$\,s$^{-1}$; including plage and flux-emergence events). The internetwork features have a diffusion coefficient of 80-90~km$^2$\,s$^{-1}$. The MBFs around the network patch in ROI-2 have a value of 130~km$^2$\,s$^{-1}$, the largest among the regions studied here. The latter implies that the random walkers sweep the largest area per unit time, compared to those moving in a preferred direction.

The diffusion coefficients have rather wide distributions with relatively large standard deviations (see Tabel ~\ref{table:stat_obs}). This is, however, not surprising, since $D$ has been shown to be dependent on spatial and temporal scales (e.g., \citealt{Abramenko2011, Giannattasio2013,Jafarzadeh2014a}). Thus, with a wide range of lifetimes of the MBFs under study, a wide range of $D$ values are obtained. The rather small FOV of the ROIs under study excludes the effect of large spatial scales on $D$.

\begin{figure}[!hp]
\centering
    \includegraphics[width=8.5cm, trim = 0 0 0 0, clip]{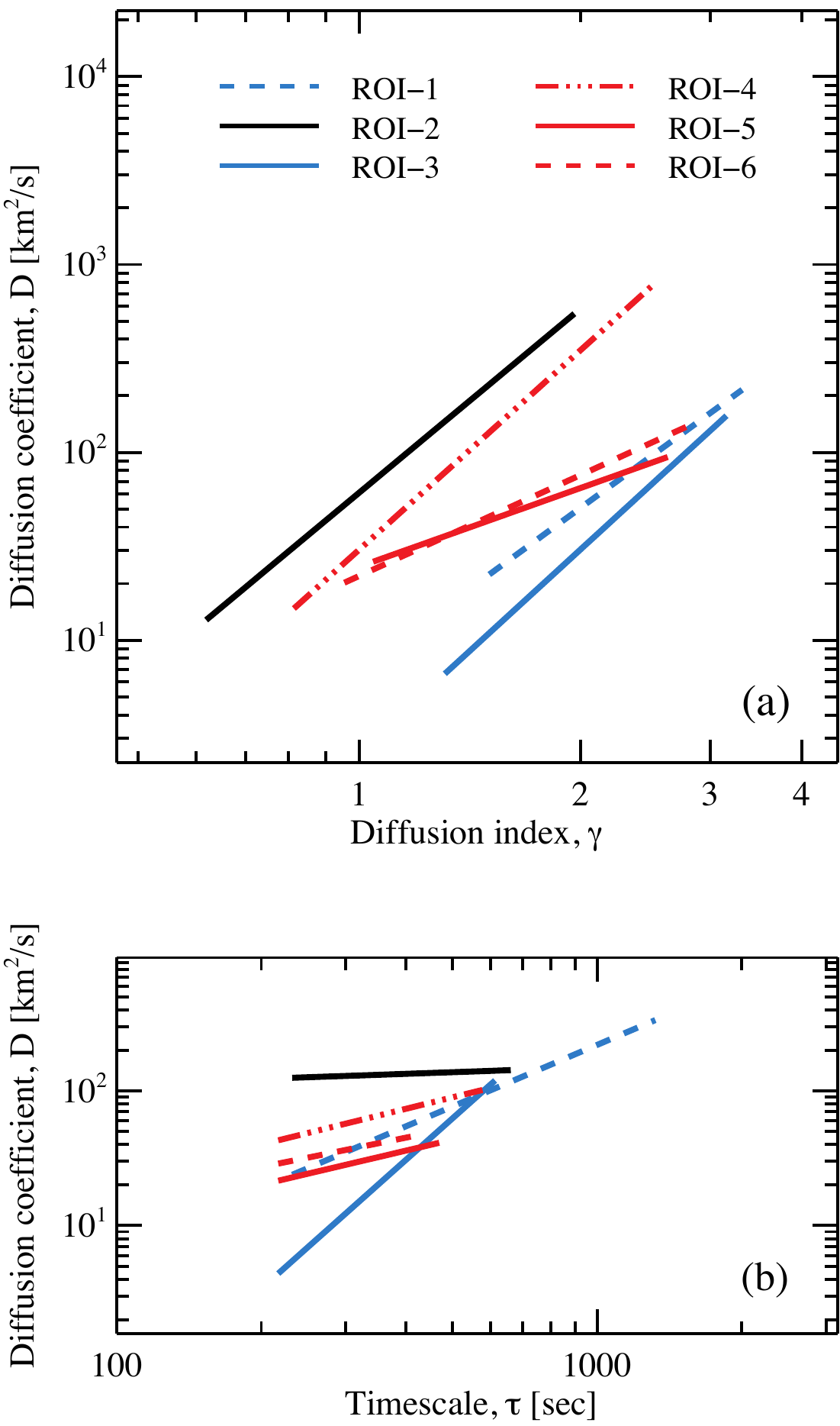}
  \caption{Log-log plots of diffusion coefficient, $D$, as a function of diffusion index, $\gamma$ (a), and as a function of timescale, $\tau$ (b) for the six regions of interest (ROI) identified by the numbers labeled in Figures~\ref{fig:obsI} and \ref{fig:obsII}. The linear fits to the data points for the various ROIs are shown with different line styles and colors introduced in panel (a).}
  \label{fig:D_gamma_tau}
\end{figure}

$D(\gamma)$ and $D(\tau)$ are plotted in Figures~\ref{fig:D_gamma_tau}(a) and \ref{fig:D_gamma_tau}(b), respectively. The different lines, introduced in panel (a), represent the best linear fits to the data points corresponding to the various ROIs. Similar ROIs, i.e., the internetwork, network, and active regions, are indicated with the colors, blue, black, and red, respectively. The diffusion coefficient of the MBFs under study is found to be directly correlated with the diffusion index, with slightly different regression slopes for the different regions. The parameter $D$ is almost independent of timescale (i.e., lifetime of the MBFs) for the random walkers, has a small dependency on timescale for MBFs observed in active regions, and is strongly correlated with $\tau$ for the internetwork MBFs (see Figure~\ref{fig:D_gamma_tau}(b)).

\begin{table*}[!thp]
\begin{center}
\caption{Diffusion Parameters of Magnetic Bright Features Observed in Various Solar Regions by {\sc Sunrise}/IMaX.}
\label{table:stat_obs}
\setlength{\tabcolsep}{1.45em}   
\renewcommand{\arraystretch}{1.6}         
\begin{tabular}{l c c c c c c c c c}
\Xhline{0.8pt}
\\[-2em]
\Xhline{0.8pt}
\\[-1.4em]
\multicolumn{2}{c}{ROI \tablenotemark{a}} & Number of & \multicolumn{3}{c}{Diffusion Index, $\gamma$} && \multicolumn{3}{c}{Diffusion Coefficient, $D$ [km$^2$\,s$^{-1}$]} \\
\cline{1-2} \cline{4-6} \cline{8-10}
Description & No. & Features ($n$) & Mean & $\sigma$ \tablenotemark{d} & $\sigma_{M}$ \tablenotemark{e} && Mean (km$^2$\,s$^{-1}$) & $\sigma$ \tablenotemark{d} (km$^2$\,s$^{-1}$) & $\sigma_{M}$ \tablenotemark{e} (km$^2$\,s$^{-1}$) \\
\Xhline{0.8pt}
\multicolumn{1}{l|}{\multirow{2}{*}{Internetwork}} & 1 & 121 & 1.9 & 0.7 & 0.06 && 79 & 170 & 15 \\
\multicolumn{1}{l|}{} & 3 & 31 & 2.0 & 0.7 & 0.13 && 92 & 211 & 38 \\
\Xhline{0.8pt}
\multicolumn{1}{l|}{Network} & 2 & 21 & 1.0 & 0.4 & 0.09 && 127 & 238 & 52 \\
\Xhline{0.8pt}
\multicolumn{1}{l|}{\multirow{3}{*}{\makecell[l]{Active\\Regions}}}  & 4 & 33 & 1.4 & 0.5 & 0.09 && 67 & 118 & 21 \\
\multicolumn{1}{l|}{} & 5 & 57 &1.5 & 0.7 & 0.09 && 40 & 121 & 16 \\
\multicolumn{1}{l|}{} & 6 & 42 & 1.4 & 0.7 & 0.11 && 24 & 81 & 12 \\
\Xhline{0.8pt}
\vspace{-6mm}
\end{tabular}
\end{center}
\hspace{1mm}
\textbf{Notes.}
\vspace{0.7mm}
\footnotetext[1]{ Region of interest, with numbers labeled in Figures~\ref{fig:obsI} and \ref{fig:obsII}.}
\footnotetext[4]{ Standard deviation of distributions.}
\footnotetext[5]{ Uncertainty in the mean values (i.e., $\sigma/\sqrt{n}$).\vspace{3mm}}
\end{table*}

\vspace{3mm}
\section{Discussion and Conclusions}
\label{conclusions}

We have presented diffusion parameters of MBFs in six areas on the solar surface (close to the disk center) with various levels of magnetic activity. These include quiet-Sun internetwork and network regions, an internetwork area in the vicinity of an active region, an area hosting flux emergence, a region consisting of plage and a few small pores, and a part of an active region including relatively large pores. We have used seeing-free observations with high spatial and temporal resolution recorded by {\sc Sunrise}/IMaX to analyze trajectories of the MBFs in the six ROIs. The seeing-free data has secured our results against the effects of, e.g., non-solar turbulence caused by variable seeing.

We found that the MBFs are super-diffusive in all regions except in the immediate vicinity of network areas. Our analysis revealed that the MBFs are transported with a nearly constant speed ($\gamma=1.9\pm0.06$ and $\gamma=2.0\pm0.13$) in both internetwork areas studied here (ROI-1 and ROI-3; see Figures~\ref{fig:obsI} and \ref{fig:obsII}), while they are best described as random walkers close to the network region ($\gamma=1.0\pm0.09$; ROI-2). Finally, the MBFs residing in a plage region or close to pores within an active region have a decelerating motion. For the latter situation, we found similar diffusion indices of $\gamma=1.4\pm0.09$, $\gamma=1.5\pm0.09$, and $\gamma=1.4\pm0.11$ for the three areas in ROI-4, ROI-5, and ROI-6, respectively.

\citet{Abramenko2011} determined diffusion indices for three different regions observed with the NST at the Big Bear Solar Observatory: coronal hole ($\gamma$=1.67), quiet-Sun ($\gamma$=1.53), and plage ($\gamma$=1.48). They provided snapshots of coronal hole and quiet-Sun areas that clearly represent an internetwork and a network region, respectively. In fact, most of the trajectories they have depicted on their ``quiet-Sun'' image overlap with network field concentrations. Thus, in a general agreement with our findings, they also found that the diffusion index decreases from the internetwork (their most quiet-Sun region) toward the more active plage areas. The value for their network magnetic elements is, however, relatively large and does not describe random walkers (or features in the sub-diffusive regime), which are expected for areas with stagnation points.

\citet{Keys2014} reported an average diffusion index of 1.2 for both quiet-Sun internetwork and active regions, which is smaller than those we obtained for regions with similar levels of magnetic flux. \citet{Yang2015} found $\gamma$ values of 1.53 and 1.79 for migration of GBPs observed with Hinode/BFI in an active region and a quiet area, respectively. Their active-region value agrees with that determined in this study.

The mean $\gamma$ value of our network MBFs is comparable with those found by, e.g., \citet{Cadavid1999} ($\gamma$=0.76--1.10, depending on timescales), \citet{Lawrence2001} ($\gamma$=1.13), and \citet{Giannattasio2014} ($\gamma$=1.08-1.27 for different scales) in network areas. Larger diffusion indices have also been reported for network regions by, e.g., \citet{Berger1998a} ($\gamma$=1.34). Our internetwork $\gamma$ value is larger than those reported in the literature. In addition to those mentioned above, examples of diffusion indices of internetwork magnetic features are $\gamma$=1 \citep{Utz2010}, $\gamma$=0.96 \citep{Manso-Sainz2011}, $\gamma$=1.59 \citep{Chitta2012}, $\gamma$=1.20--1.34 (for different length scales; \citealt{Giannattasio2013}), $\gamma$=1.69 (Paper~I), and $\gamma$=1.44 \citep{Giannattasio2014}. We note that the $\gamma$ value in Paper~I was determined from migration of magnetic bright points in both internetwork and vicinity of network areas. The mixed contributions of both internetwork and network regions led to the intermediate $\gamma$ value of 1.69, that is closer to the internetwork one in the present study, as the internetwork covered a larger fraction of the area in the image sequences employed in Paper~I.

We also found a significant scatter of $\gamma$ values obtained from individual MBFs within a single type of region. We interpret the wide spread in parameters we obtained as the effect of realization noise (i.e., short lifetimes of the features and/or of the image sequences; based on a discussion in Paper~I).

The coefficients of turbulent diffusion are found to depend on the level of magnetic activity in the ROI under study. Thus we obtained the smallest value ($D=25$~km$^2$\,s$^{-1}$) for ROI-6, which includes large pores. The values of $D=40$~km$^2$\,s$^{-1}$ and $D=70$~km$^2$\,s$^{-1}$ are obtained for the two other parts of active regions with pores/plages and plages/flux emergence in ROI-5 and ROI-4, respectively. The internetwork areas found to host MBFs with $D=80$~km$^2$\,s$^{-1}$ and $D=90$~km$^2$\,s$^{-1}$ in ROI-1 and ROI-3, respectively. The diffusion coefficient in the network region (ROI-2) was, however, found to be the largest value ($D=130$~km$^2$\,s$^{-1}$) among all regions considered in this work. This indicates that the random walkers sweep larger areas in time by randomly moving around compared to super-diffusive features that migrate with a preferred direction.

A wide range of diffusion coefficients have been reported in the literature. For recent reviews of some of these values we refer the reader to Paper~I and \citet{Yang2015}. Our study, in particular, agrees with that of found by \citet{Yang2015} who also obtained an anti-correlation between $D$ and the level of magnetic flux. They reported $D=78\pm29$ for active regions and $D=130\pm54$ for a quiet-Sun area. Most of their quiet-Sun features are located at, or close to, network regions.

We found that $D$ is nearly independent of timescale for MBFs observed in the vicinity of network area (i.e., random walkers; ROI-2), whereas $D$ has a direct correlation with $\tau$ for all the other (super-diffusive) MBFs under study. The increase of $D$ with timescale is faster for internetwork MBFs (seen in ROI-1 and ROI-3) compared to those observed in active regions (i.e., in ROI-4, ROI-5, and ROI-6). Due to relatively small ROIs in our study, we did not inspect the correlations between values of $D$ and length scales. We also found a direct relationship between $D$ and $\gamma$ for all the six ROIs.

All the diffusion coefficients we obtained here, and in particular those for the quiet-Sun internetwork, are smaller than that found in Paper~I for internetwork Ca~{\sc ii}~H magnetic bright points (270~km$^2$\,s$^{-1}$). The latter measurement was based on observations in the Ca~{\sc ii}~H filter from {\sc Sunrise}/SuFI \citep{Gandorfer2011}, sampling a higher atmospheric layer (roughly corresponding to the temperature minimum) than the data we employed in this study. Because of a decrease in mass density with height, we expect that magnetic elements (as cross sections of flux tubes) sweep larger areas in time at the heights sampled by the SuFI Ca~{\sc ii}~H channels than those in the lower photosphere. 
This is also suggested by the 2D MHD simulations of \citet{Steiner1998}, who show that the upper parts of flux sheets swing back and forth much more than their lower layers when plied by the surrounding convection.
Alternatively, the difference between the diffusion coefficient obtained for the Ca~{\sc ii}~H magnetic bright points in Paper~I and those found here can be the result of the anti-correlation between $D$ and size of magnetic features, as was previously shown by \citet{Schrijver1996}. In Paper~I, only magnetic features smaller than 0.3~arcsec were analyzed, whereas no size threshold was applied to the MBFs in this study.

The difference between the various values of the diffusion parameters, and in particular $D$, reported in the literature (including those we obtained in the present study) can be due to a number of factors. These include properties of different datasets with various spatial and/or temporal resolutions, the method with which mean values are calculated, time and length scales, the atmospheric height at which the magnetic elements are sampled, size and magnetic strength of features, and the level of magnetic flux of the region hosting magnetic elements.
Diffusion analysis of synthesized images from 3D radiative MHD simulations of regions with different amounts of magnetic flux, different resolution, and sampling various atmospheric layers can provide us with a better understanding of such variations.

\acknowledgements
The German contribution to {\sc Sunrise} and its reflight was funded by the Max Planck Foundation, the Strategic Innovations Fund of the President of the Max Planck Society (MPG), DLR, and private donations by supporting members of the Max Planck Society, which is gratefully acknowledged. The Spanish contribution was funded by the Ministerio de Econom\'{\i}a y Competitividad under Projects ESP2013-47349-C6 and ESP2014-56169-C6, partially using European FEDER funds. The HAO contribution was partly funded through NASA grant number NNX13AE95G. This work was partly supported by the BK21 plus program through the National Research Foundation (NRF) funded by the Ministry of Education of Korea. S.J. receives support from the Research Council of Norway. The National Solar Observatory (NSO) is operated by the Association of Universities for Research in Astronomy (AURA) Inc. under a cooperative agreement with the National Science Foundation.

\vspace{5mm}
\bibliographystyle{aa}
\bibliography{Shahin}

\end{document}